\documentclass[graybox]{svmult}

\usepackage{mathptmx}       
\usepackage{helvet}         
\usepackage{courier}        
\usepackage{type1cm}        
\usepackage{makeidx}         
\usepackage{graphicx}        
\usepackage{multicol}        
\usepackage[bottom]{footmisc}
\def\actaa{\ref@jnl{Acta Astron.}}      
\def\araa{\ref@jnl{ARA\&A}}             
\def\apjl{\ref@jnl{ApJ}}                
\def\ao{\ref@jnl{Appl.~Opt.}}           
\def\aap{\ref@jnl{A\&A}}                
\def\aapr{\ref@jnl{A\&A~Rev.}}          
\def\aaps{\ref@jnl{A\&AS}}              
\def\azh{\ref@jnl{AZh}}                 
\def\baas{\ref@jnl{BAAS}}               

\makeindex             

\begin{document}

\title*{A SAURON study of dwarf elliptical galaxies in the Virgo Cluster: kinematics and stellar populations}
\titlerunning{ A SAURON study of dEs in the Virgo Cluster: kinematics and stellar populations}
\author{Agnieszka Ry\'{s} and Jes\'{u}s Falc\'{o}n-Barroso}
\institute{Agnieszka Ry\'{s}, Jes\'{u}s Falc\'{o}n-Barroso \at Instituto de Astrof\'{i}sica de Canarias, C/V\'{i}a L\'{a}ctea, s/n, La Laguna, Tenerife, Spain\\
Depto. Astrof\'{i}sica, Universidad de La Laguna (ULL), E-38206 La Laguna, Tenerife, Spain\\  \email{arys@iac.es, jfalcon@iac.es}
}
%
%
\maketitle

\abstract{Dwarf elliptical galaxies (dEs) are the most common galaxy type in nearby galaxy clusters; even so, many of their basic properties have yet to be quantified. Here we present the results of our study of 4 Virgo dwarf ellipticals obtained with the SAURON integral field unit on the William Herschel Telescope (La Palma, Spain). While traditional long-slit observations are likely to miss more complicated kinematic features, with SAURON we are able to study both kinematics and stellar populations in two dimensions, obtaining a much more detailed view of the mass distribution and star formation histories. What is visible even in such a small sample is that dEs are not a uniform group, not only morphologically, but also as far as their kinematic and stellar population properties are concerned. We find the presence of substructures, varying degrees of flattening and of rotation, as well as differences in age and metallicity gradients. We confirm that two of our galaxies are significantly flattened, yet non-rotating objects, which makes them likely triaxial systems. The comparison between the dwarf and the giant groups shows that dEs could be a low-mass extension of Es in the sense that they do seem to follow the same trends with mass. However, dEs as \textit{progenitors} of Es seem less likely as we have seen that dEs have much lower abundance ratios.}

\section{Introduction}
Dwarf elliptical galaxies are the most common galaxy type in the nearby universe, accounting, for about 75\% of all objects in the Virgo cluster (e.g. \cite{trentham:2002}).  They display a variety of shapes (from round to significantly flattened) and (sub)structures (disk features, bars or spiral arms). These different properties may suggest that more than just one mechanism was involved in their formation. Even though dEs may provide important clues on the main processes involved in galaxy assembly and evolution, many of their properties are still unknown and no present formation scenario(s) can fully explain observational characteristics of all dEs subclasses. A detailed analysis of the structure, internal kinematics and stellar populations of dEs is, therefore, essential to understanding the evolutionary properties of this class of objects.


%
%
%
%

\section{Observations} 
\begin{figure*}
\centering
\includegraphics[width=0.24\textwidth]{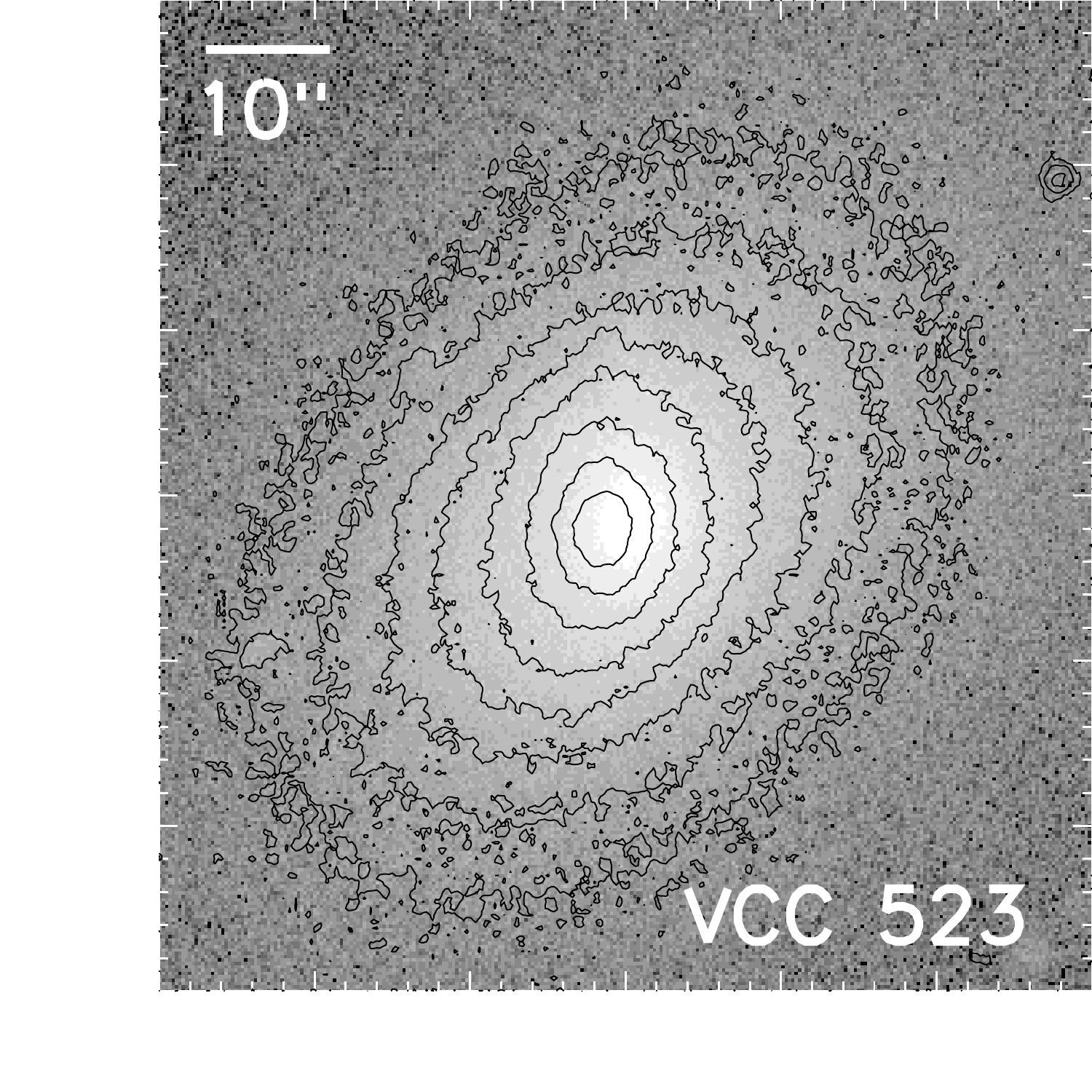}
\includegraphics[width=0.24\textwidth]{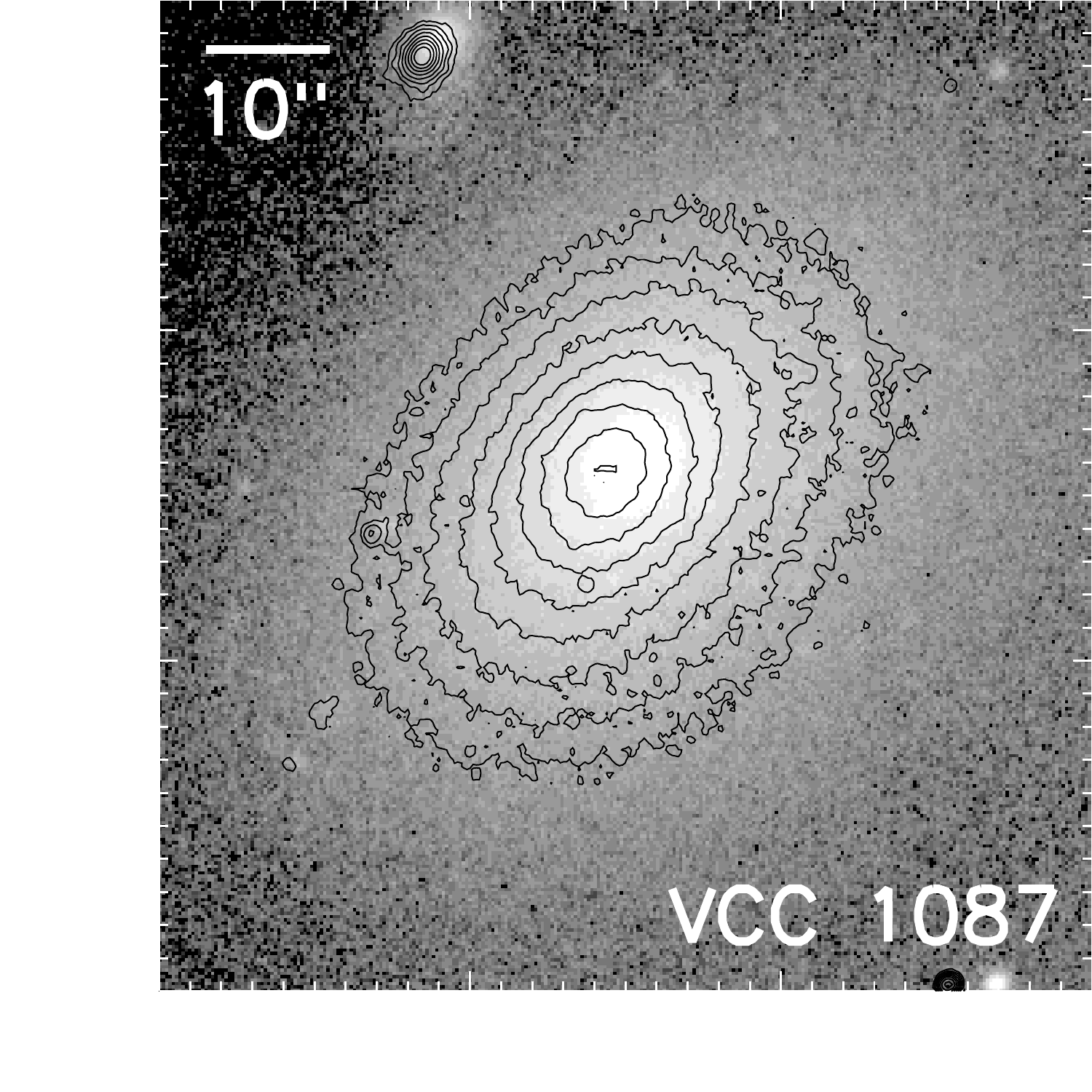}
\includegraphics[width=0.24\textwidth]{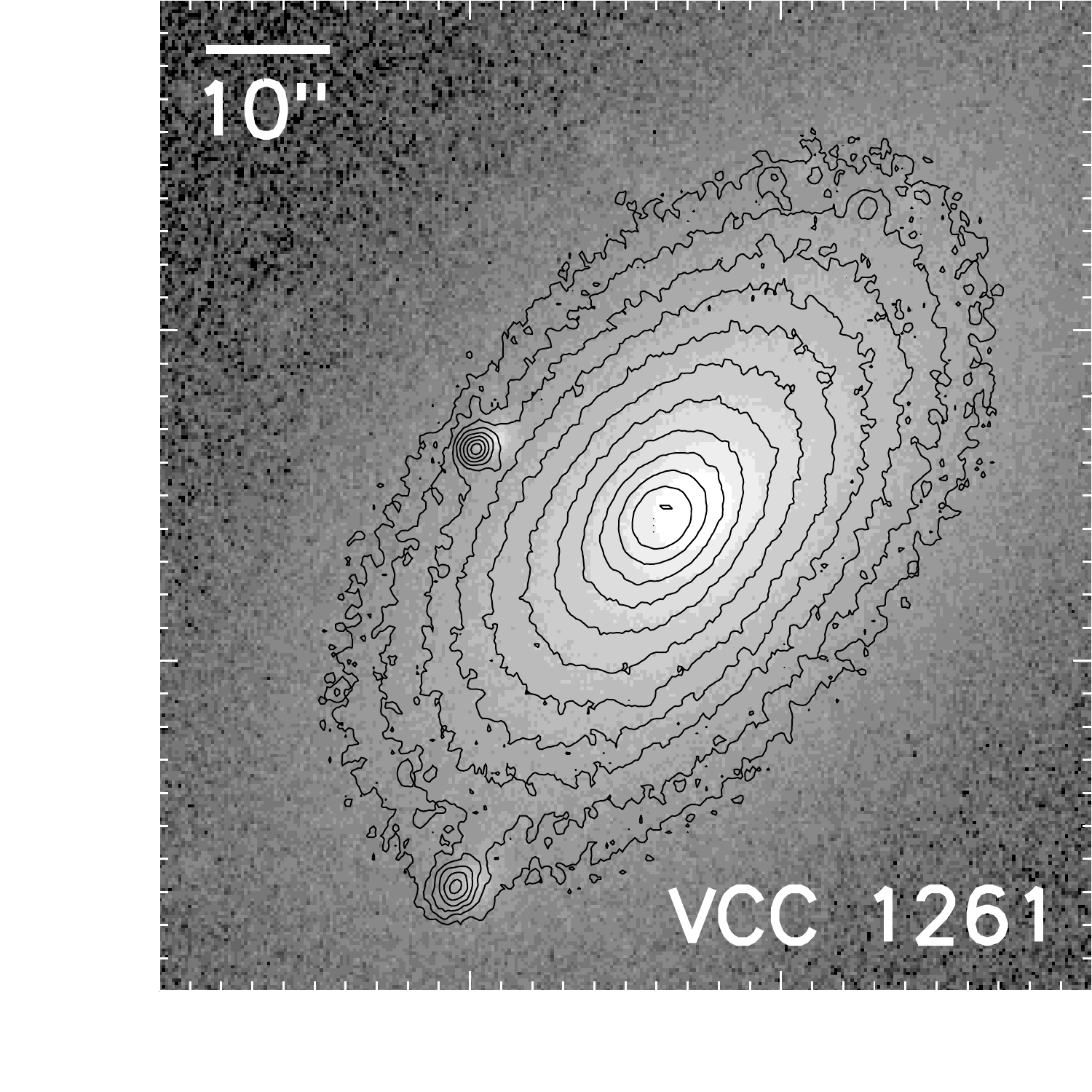}
\includegraphics[width=0.24\textwidth]{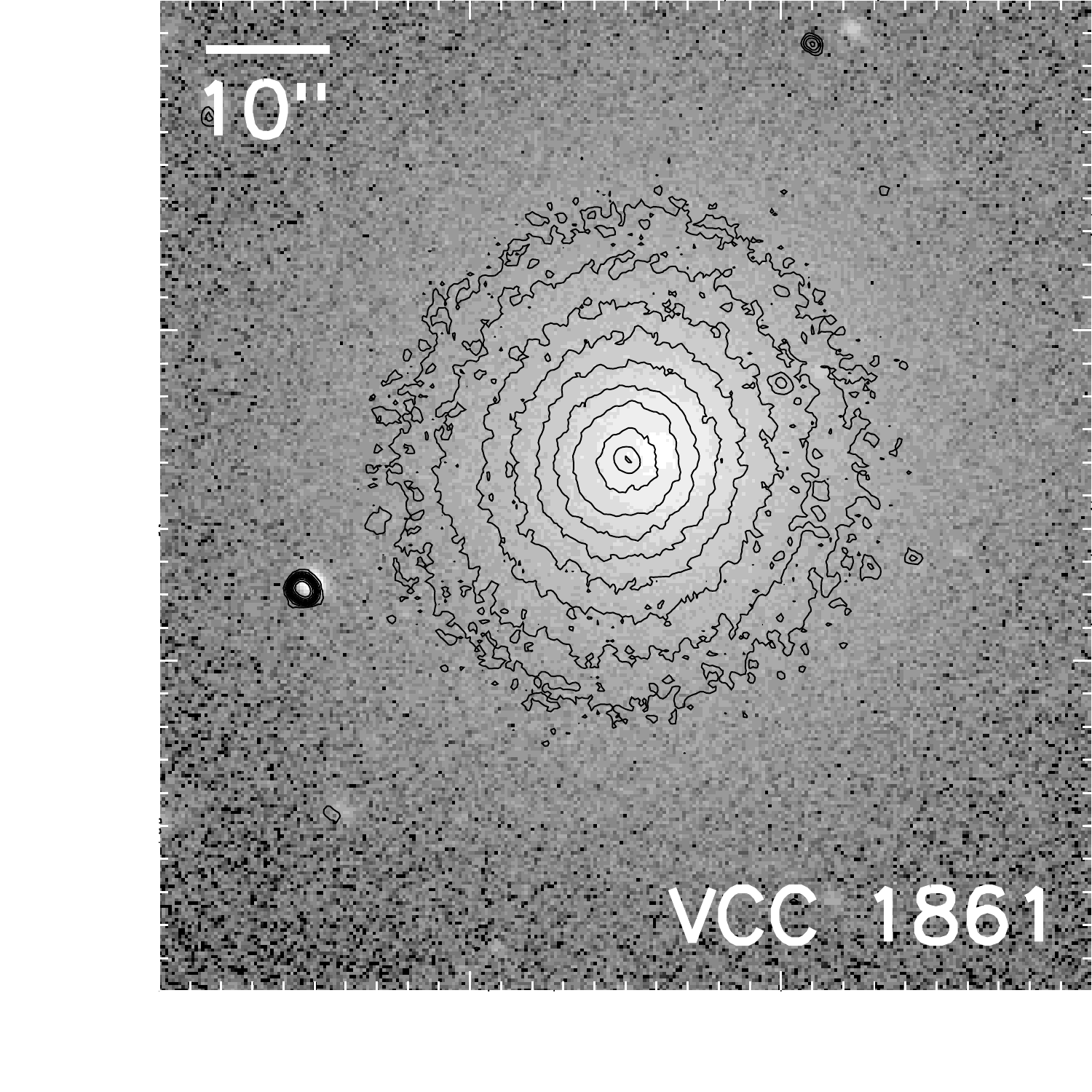}

\caption{ 75''x75'' cutouts of ACAM images for all 4 Virgo dEs with intensity contours overplotted.}
\label{acam}       
\end{figure*}

The sample consists of four Virgo bright, nucleated dEs, drawn from the Virgo catalog of \cite{lisker:2006}. We observed the targets  during 3 nights in January 2010 using the SAURON instrument mounted on the 4.2m William Herschel Telescope at the Observatorio del Roque de los Muchachos in La Palma, Spain. The average exposure time was 5h per galaxy. Additionaly, we used the WHT auxiliary camera ACAM to obtain deep images (300s) for each object (see Figure~\ref{acam}). 

\section{Stellar kinematics}
We used the penalized pixel fitting (pPXF) method of \cite{cappellari:2004} to derive stellar absorption line kinematics by directly fitting the spectra in the pixel space.We used absorption spectra taken from the single-burst stellar population models of \cite{vazdekis:1999}.

\begin{figure*}[!t]
\centering
\includegraphics[width=1.00\textwidth]{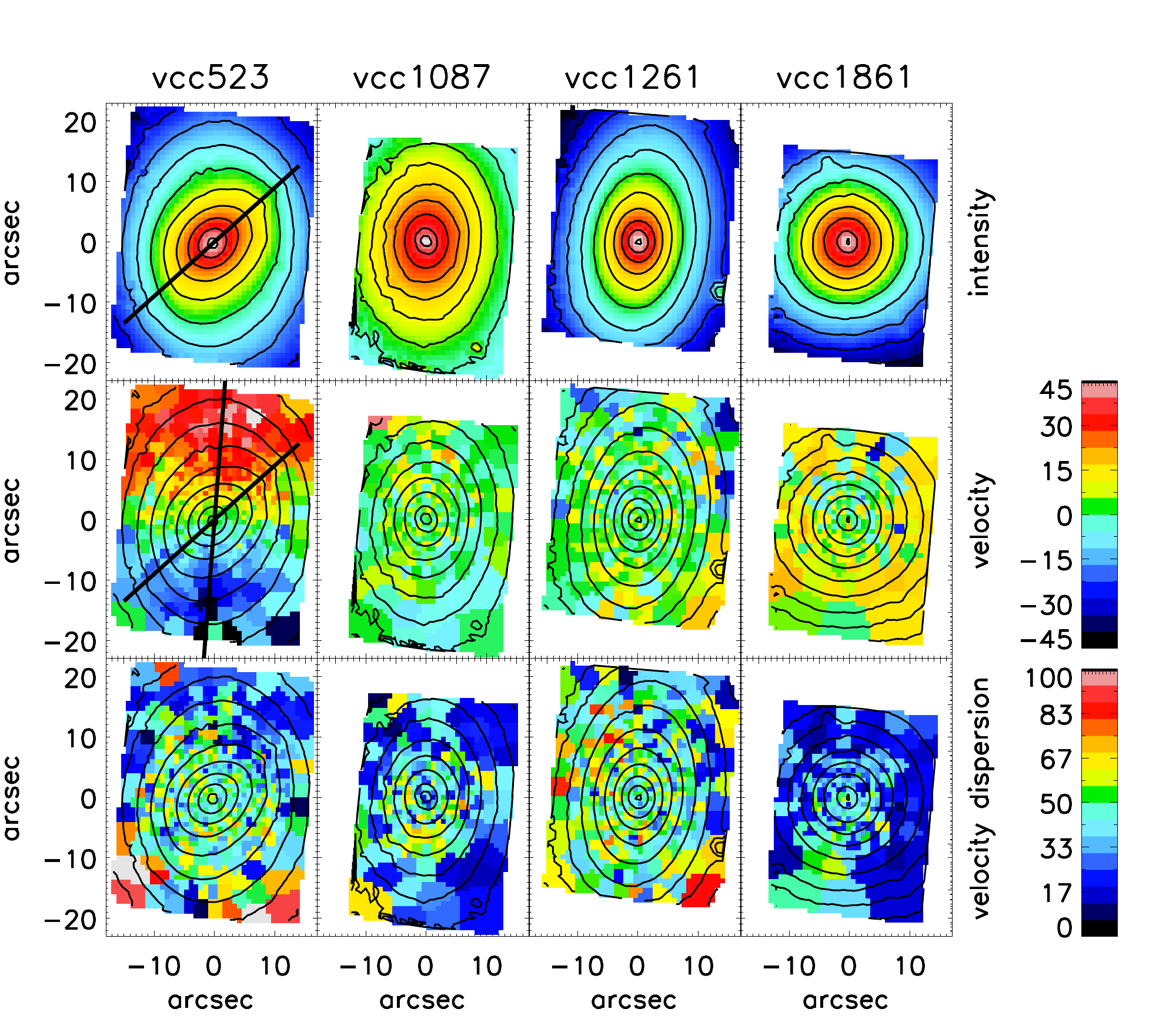}
\caption{Reconstructed instensity (top), stellar velocity $V$ (middle), and velocity dispersion $\sigma$ (bottom) maps. The first column shows VCC 523, the only clear rotator in the sample, with the photometric and kinematic major axes overplotted on the intensity and velocity maps. The two following columns show VCC 1087 and 1261, the two visibly flattened galaxies that display no rotation.}
\label{map_all}       
\end{figure*}

The obtained intensity, stellar velocity $V$ and velocity dispersion $\sigma$ maps for the four objects are shown in Figure~\ref{map_all}. The first column shows VCC 523, the only clear rotator in the sample, known to  host a bar, which is evidenced here by the misalignment between the (central) photometric and the kinematic major axes, overplotted on the intensity and the velocity map. A similar kind of substructure has been found in the SAURON project sample in at least 5 giant ellipticals (see \cite{emsellem:2004}).

VCC 1087 and 1261 (second and third column, respectively) are both visibly flattened objects, yet we do not see any rotation in our maps (the same result was previously reported by \cite{chilingarian:2009} and \cite{toloba:2010} based on long-slit data). One simple explanation could be that we do not have sufficient radial coverage but another possibility is that they are triaxial systems. Interestingly, in a recent paper \cite{beasley:2009} find that globular clusters in these two systems (found at 4-7 effective radii) show a high level of rotation. Since these objects are fairly close to the cluster's center, our result is in agreement with that of \cite{toloba:2010} who find more rotating objects at larger distances. 

When compared with the SAURON giant ellipticals sample (see Figure~9 in \cite{emsellem:2007}), we can see that the vast majority of objects with the same flattening have high angular momentum (i.e. -- are fast rotators) and only very few have low rotation velocities. In contrast, 2 out of 3 of our highly flattened objects show virtually no rotation.

An important finding is the existence of slow rotators in the dE family that \textit{kinematically} resemble the non-rotating galaxies of the E class. More specifically, our dEs compare to the three highest-mass Es from the SAURON sample, i.e. those that do not have a kinematically-decoupled core (KDC). \textit{Photometrically}, our dEs compare to the flattened Es (i.e. have similar flattening); \textit{kinematically} they resemble each other in the outer parts, while we are missing the KDCs.

The question arises: should we be able to detect the presence of KDCs in dwarfs? This depends on their size: assuming the ``worst-case'' scenario for the size of the potential KDCs in dwarfs, i.e. taking the largest effective radius (R$_{eff}$) / KDC size ratio for Es (the values from \cite{mcdermid:2006}) and combining it with the smallest R$_{eff}$ for dwarfs, we find that we could expect KDCs of the size of 3-4 arcsec. This is at the limits of what we could resolve with our instrument, considering the seeing and sampling effects. Confirming the presence of KDCs in dwarfs would be a major argument in favor of them being scaled down versions of giant ellipticals.

\section{Stellar population parameters}


Our wavelength range allows us to measure four indices: H$\beta$, Mg$b$, Fe5015 and Fe5270, of which H$\beta$ is the  age-sensitive index and the rest are metallicity indicators. However, these metallicity indices \textit{are}  affected by age and abundance changes, which is why we use abundance-ratio insensitive index  combination [MgFe50] of \cite{kuntschner:2010} in our analysis\footnote{$[\mathrm{MgFe50}] = \frac{0.69 \cdot \mathrm{Mg}b + \mathrm{Fe}5015}{2}$}. We also use a new optimized H$\beta_o$ index, defined in \cite{cervantes:2009}, which is less dependent on metallicity than the traditional Lick H$\beta$ index.

\begin{figure*}[!t]
\centering
\includegraphics[width=1.00\textwidth]{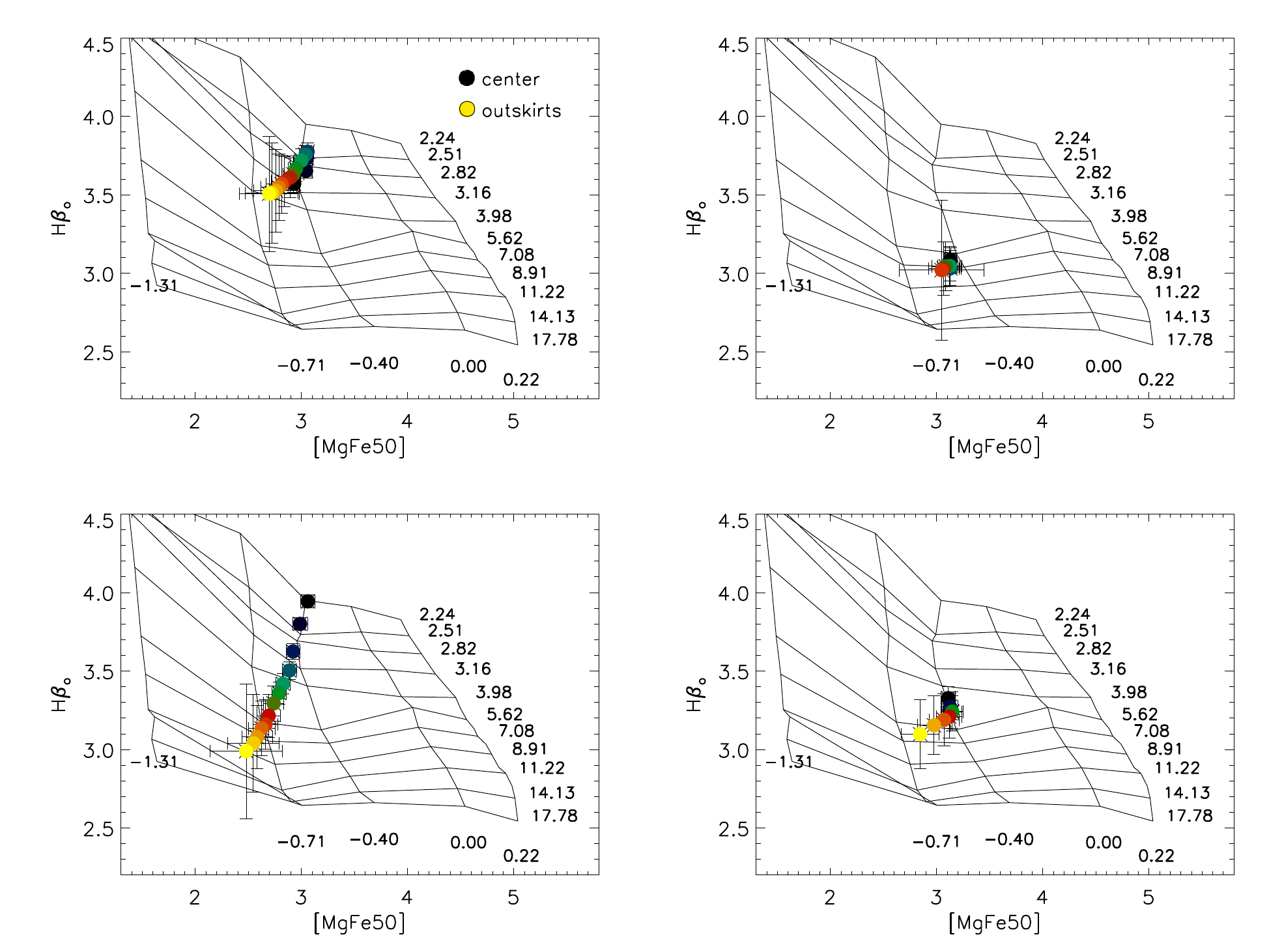}
\caption{H$\beta_o$ versus [MgFe50] for VCC 523 (upper left), VCC 1087 (upper right), VCC 1261 (lower left) and VCC 1861 (lower right). While all the galaxies get older and more metal-poor in the outer parts, the actual gradients vary significantly among them.}
\label{i-i}  
\end{figure*}

\begin{figure*}
\centering
\includegraphics[width=0.52\textwidth]{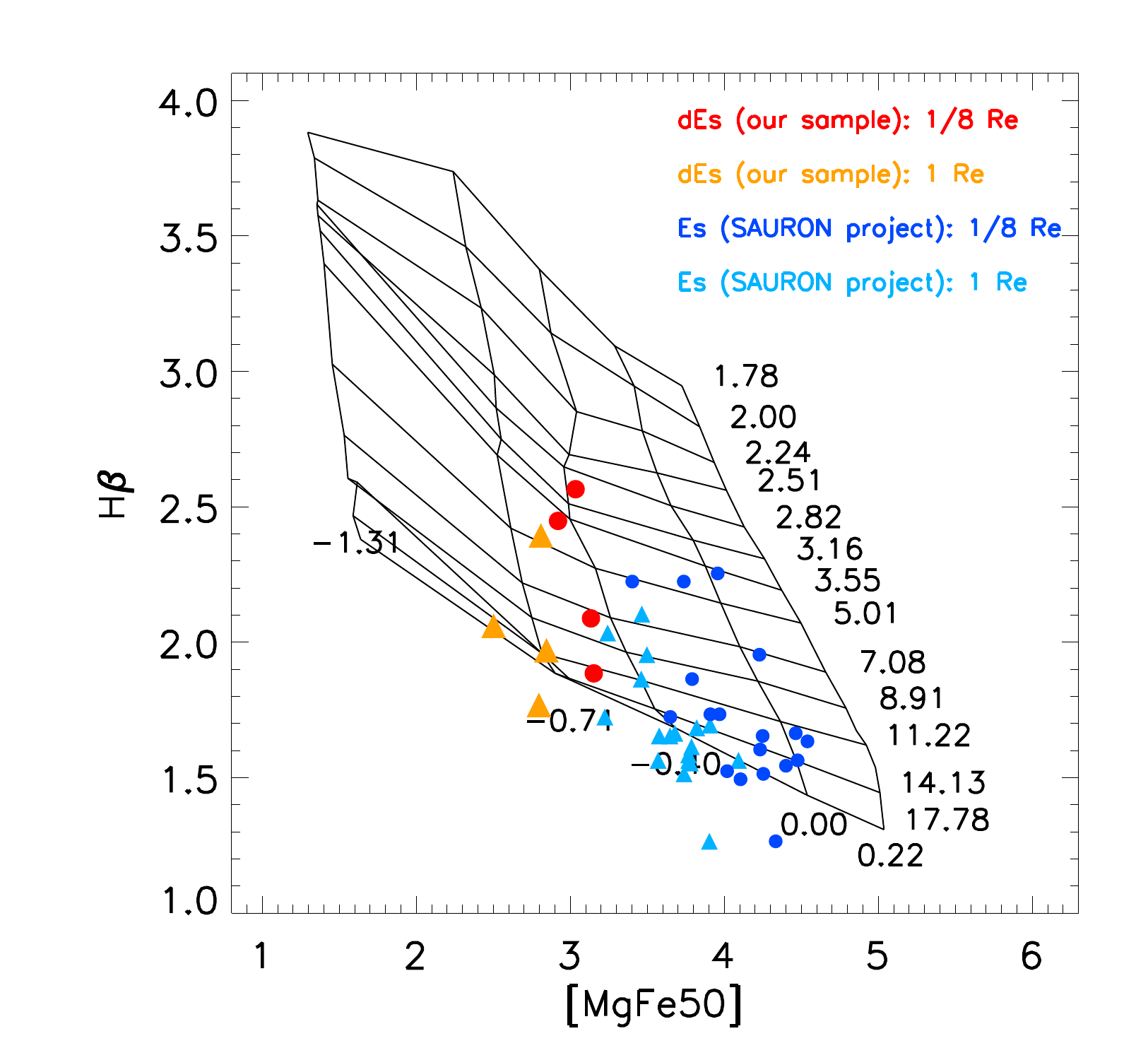}
\caption{H$\beta$ versus [MgFe50], values for dE and Es  integrated within 1/8 R$_{eff}$ and 1 R$_{eff}$. As expected, the galaxies get older and more metal-poor as we move away from the center. Also, dEs are on average younger and more metal-poor than Es.}
\label{i-comp}       
\end{figure*}

In Figure~\ref{i-i} we show index profiles obtained by averaging over ellipses at different galactocentric distances. The trends are similar in that all galaxies, as expected, get older and more metal poor as we move away from the center, but the actual gradients, especially in age, vary significantly.

We compared our results with those for the SAURON project sample of 17 Virgo Es, using values integrated within 1/8 and 1 $R_{eff}$ whose values we took from \cite{toloba:2010} (Figure~\ref{i-comp}).  Since our data do not go as far as one full $R_{eff}$, aperture corrections were applied following the method of \cite{kuntschner:2006}. As expected  (see, e.g. \cite{michielsen:2008}), the dwarfs are younger and more metal poor than their giant counterparts.

\begin{figure*}[!t]
\centering
\includegraphics[width=1.00\textwidth]{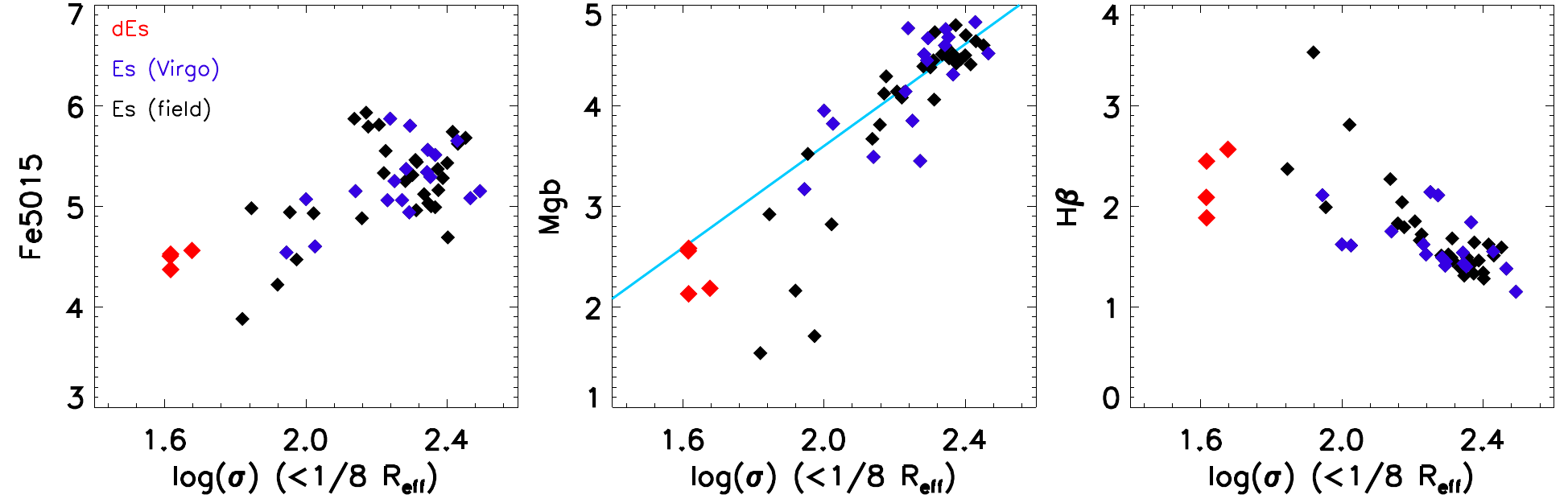}
\caption{H$\beta_o$, Mg$b$ and Fe5015 as a function of $\sigma$ for our sample of 4 dEs (red) and giant Es from the SAURON project:  Virgo (blue) and field (black). }
\label{i-sig}       
\end{figure*}

To better determine how the dwarf and the giant families compare, we plotted various quantities as a function of $\sigma_e$ ($\sigma$ integrated within 1 R$_{eff}$) and dynamical mass M$_{dyn}$ \footnote{M$_{dyn}=5.0\cdot R_e \cdot \sigma_e^2/G$}. (Figure~\ref{i-sig}). The middle plot shows the well known Mg$b$-$\sigma$ relation, any deviations from which are indicative of the presence of younger stellar populations.  Our objects seem to follow the same trend as the old Es, even though we do have young populations in dEs - this fact does not seem to be able to influence the trend.

The H$\beta$ plot (right) shows dEs as being located below the general trend. This is likely due to the SAURON sample selection, which is distance and magnitude limited -- there is significant  scatter present in the low-$\sigma$ area in more complete samples of, e.g. \cite{thomas:2002}.

\begin{figure*}[!t]
\centering
\includegraphics[width=1.00\textwidth]{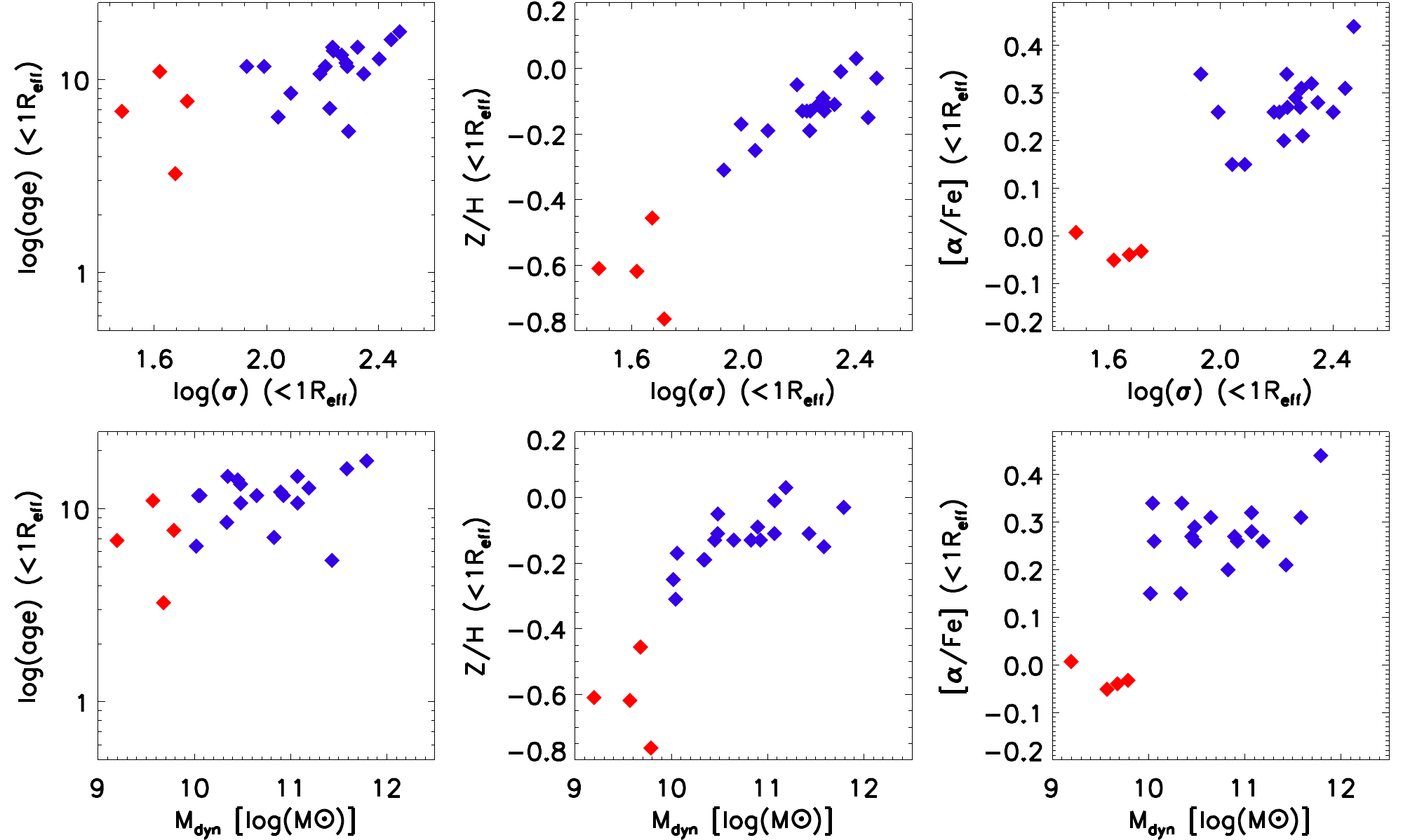}
\caption{Age (left), metallicity (middle) and abundance (right) as a function of velocity dispersion $\sigma$ (upper panel) and dynamical mass M$_{dyn}$ (lower panel) for our sample of 4 dEs (red) and Virgo giant Es from the SAURON project (blue). }
\label{age-sig}       
\end{figure*}

In Figure~\ref{age-sig} we show the values obtained for age, Z and [$\alpha$/Fe] as a function of velocity dispersion $\sigma$ (upper panel) and dynamical mass M$_{dyn}$. Our dEs fall on the extension of the best fit found for giant ellipticals. However, the apparently tight age-$\sigma$ relation could, again, be the result of the selection effects present in the SAURON sample. The fact that more complete samples show increased scatter with decreasing mass can be explained in the following way. When a rejuvenating event takes place in a massive galaxy, its overall impact on the galaxy's characteristics is not as significant as in the case of dEs where a star-formation event affects a much larger part of the galaxy - that's where the scatter at lower masses comes from.

Another important point is that dEs have much lower abundance ratios than Es, which has important implications for theories of galaxy formation. Clearly, it would be difficult to think of a formation scenario that would explain how low-abundant dwarfs could be progenitors of  highly-abundant giants. It is therefore unlikely for the dwarfs to be ``building blocks'' in the traditional sense.

\section{Summary and conclusions}

We have presented the results of our SAURON study of four bright nucleated dwarf ellipticals from the Virgo Cluster. We have obtained reliable stellar velocity and velocity dispersion maps. We have also measured line-strength indices that allowed us to estimate ages and metallicities and compare the trends with those obtained for giant ellipticals from the SAURON sample. 

We find one rotating galaxy with misaligned photometric and kinematic major axes, indicating the presence of a bar. Two out of three of our objects that do not appear to rotate are significantly flattened, we speculate that they might be triaxial systems. We calculate line-strength indices and compare them with the predictions from the new MILES models \cite{vazdekis:2010}. The index-index diagrams show the presence of age and metallicity gradients different for each galaxy. 

The comparison with the Virgo giant ellipticals (Es) sample from the SAURON project \cite{kuntschner:2010} reveals that dEs have lower metallicity and are on average younger than Es. Both types seem to follow the same trends when index values as well as age, Z, and abundances are plotted as a function of velocity dispersion and dynamical mass. We conclude that dEs can, in principle, be a low-mass extension of Es, given that they seem to follow the same trends with mass. However, dEs as progenitors of Es seem less likely since the former have much lower abundance ratios than the latter. This argument is consistent with detailed photometric studies (e.g. \cite{kormendy:2009} and references therein).



\section{Future work}
We will expand our sample during the observations scheduled for spring 2011.  In the meantime, further analysis of the data will include full spectrum fitting to obtain a more detailed view on the star formation histories of the galaxies (initial results show agreement with the age and Z values obtained from measuring indices). We will also construct dynamical models for our objects. The combination of the results from the ancillary imaging with those from the models will enable us to discuss the level of the velocity anisotropy and the orbital structure in these systems and constrain their mass distribution, including any possible contribution from dark matter.

 \bibliographystyle{spmpsci}

\bibliography{biblio} 

\end{document}